\documentclass[aps,prd,preprint,superscriptaddress,showpacs,floatfix,preprintnumbers,nofootinbib,
a4paper]{revtex4-1}
\pdfoutput=1
\usepackage{hyperref}
\usepackage{color}
\usepackage{graphicx}
\usepackage{textcomp}
\usepackage{subfigure}
\usepackage{amsmath,amssymb}
\usepackage{enumerate}
\newcommand{\mathsym}[1]{{}}

\def\lsim{\:\raisebox{-1.1ex}{$\stackrel{\textstyle<}{\sim}$}\:}

\newcommand{\ba}{\begin{array}}
\newcommand{\ea}{\end{array}}
\newcommand{\be}{\begin{equation}}
\newcommand{\ee}{\end{equation}}
\newcommand{\beqa}{\begin{eqnarray}}
\newcommand{\eeqa}{\end{eqnarray}}

\newcommand{\AddrPD}{Dipartimento  di Fisica e Astronomia `G. Galilei', Universit\`a di Padova, Via
Marzolo 8, I-35131 Padua, Italy}
\newcommand{\AddrINFN}{Istituto Nazionale di Fisica Nucleare, Sezione di Padova, I-35131 Padua,
Italy \vspace*{0.5cm}}

  \hypersetup{
   pdfstartview=FitV,
   colorlinks=true,
   linkcolor=blue,
   citecolor=blue,
   filecolor=black,
   urlcolor=blue}

\begin{document}
\vspace*{1cm}
\title{Order and Anarchy hand in hand in 5D SO(10)}

\author{Ferruccio Feruglio}
\email{feruglio@pd.infn.it}
\affiliation{\AddrPD}
\affiliation{\AddrINFN}
\author{Ketan M. Patel}
\email{ketan.patel@pd.infn.it}
\affiliation{\AddrINFN}
\author{Denise Vicino}
\email{denise.vicino@pd.infn.it}
\affiliation{\AddrPD}
\affiliation{\AddrINFN}

\bigskip


\begin{abstract}
\vspace*{0.2cm}
We update a five-dimensional SO(10) grand unified model of fermion masses and mixing
angles originally proposed by Kitano and Li. In our setup Yukawa couplings are anarchical and quark
and lepton sectors are diversified by the profiles of the fermion zero modes in the extra
dimension. The breaking of SO(10) down to SU(5)$\times$U(1)$_{\rm X}$ provides the key parameter
that distinguishes the profiles of the different SU(5) components inside the same ${\bf 16}$
representation. With respect to the original version of the model,  we extend the Higgs sector to explicitly solve the doublet-triplet splitting problem through the missing partner mechanism and we perform a fit to an idealized set of data. By scanning the Yukawa couplings of the model we find that, for large $\tan\beta$, both normal and inverted ordered neutrino spectrum can be accommodated. However, while the case of inverted order requires a severe fine tuning of the Yukawa parameters,  the normal ordering is compatible with an anarchical distribution of Yukawa couplings. Thus, in a natural portion of the parameter space, the model predicts a normal ordered neutrino
spectrum, the lightest neutrino mass below $5$ meV, and $|m_{\beta\beta}|$ in the range 0.1-5 meV.  No particular preference is found for the Dirac CP phase in the lepton sector while the right-handed neutrino
masses are too small to explain the baryon asymmetry of the universe through thermal leptogenesis.
\end{abstract}

\maketitle

\section{Introduction}
\label{introduction}
It is a matter of fact that the particle content of the Standard Model (SM) and its gauge group
perfectly fit into an SO(10) grand unified theory (GUT) \cite{Georgi:1974sy,*Fritzsch:1974nn}, with
fermions of each generation contained in a spinorial representation of the grand unified group. The
extra component of the {\bf 16} is a right-handed neutrino that nicely allows a description of the
light neutrino masses in terms of the seesaw mechanism
\cite{Minkowski:1977sc,*Yanagida:1979as,*GellMann:1980vs,*Glashow:1979nm,Lazarides:1980nt,
*Schechter:1980gr,*Mohapatra:1980yp} and provides the ingredients for successful
leptogenesis \cite{Fukugita:1986hr}. The supersymmetric (SUSY) version of the model can also
accommodate gauge coupling unification without intermediate
scales \cite{Giunti:1991ta,*Amaldi:1991cn,*Langacker:1991an}. Despite these promising features, the
description of fermion masses and mixing angles in an SO(10) model is as complicated as in the SM and
no advantage seems to come from the grand unified picture in this respect.

The product of two {\bf 16} decomposes into the sum ${\bf 10}+{\bf 120}+{\bf 126}$ and any
combination of these representations can enter the Yukawa interaction terms. A minimal model, where
only one copy of the {\bf 10} is included, is completely unrealistic. It would describe equal masses
for quarks and leptons, up to an overall factor distinguishing $\pm 1/2$ components of the weak
isospin doublets, and no mixing. It is intriguing that, on the contrary, a minimal SU(5) GUT with
Higgses in $5$ and $\bar 5$ representations comes very close to the real world, the only wrong
prediction being the exact equality between the charged lepton and down quark masses, which needs
corrections of order one. Non-minimal renormalizable models, where various combinations of Higgses
in ${\bf 10}$, ${\bf 120}$ and $\overline{\bf 126}$ representations are introduced, have been shown
to fit the fermion mass data well
\cite{Babu:1992ia,*Oda:1998na,*Bertolini:2004eq,*Babu:2005ia,*Bertolini:2005qb,Bertolini:2006pe,
*Grimus:2006rk,*Altarelli:2010at,*Joshipura:2011nn,*Dueck:2013gca,*Altarelli:2013aqa,Aulakh:2008sn,
*Aulakh:2011at}. When the $\overline{\bf 126}$ representation is included, light Majorana neutrinos
can be described by a seesaw mechanism either of type I
\cite{Minkowski:1977sc,*Yanagida:1979as,*GellMann:1980vs,*Glashow:1979nm} or of type II
\cite{Lazarides:1980nt,*Schechter:1980gr,*Mohapatra:1980yp} or by a combination of both. In these
models there is no qualitative difference with respect to the SM. The number of free parameters in
the flavor sector is very large so that no predictions are available and the best fit parameters
span several orders of magnitude, much as in the SM.

There are several SO(10) models, in both renormalizable
\cite{Clark:1982ai,*Aulakh:1982sw,*Aulakh:2003kg,*Bajc:2008dc,*Melfo:2010gf,Aulakh:2008sn,
*Aulakh:2011at,Grimus:2006bb,*Grimus:2008tm,*Joshipura:2009tg,Matsuda:2000zp,*Matsuda:2001bg,*Fukuyama:2002ch,*Fukuyama:2004xs} and
non-renormalizable
\cite{Albright:1998vf,*Babu:1998wi,Dermisek:2005ij,*Dermisek:2006dc,*Barbieri:1996ww,
*Albright:2000dk,*Albright:2001uh,*Ji:2005zk,*Alciati:2006sw,*Hagedorn:2008bc,King:2006me,
*King:2009mk,*King:2009tj,*Dutta:2009bj,*Patel:2010hr,*BhupalDev:2011gi,*BhupalDev:2012nm} versions (see
\cite{Chen:2003zv,*Fukuyama:2012rw} for reviews), where new ingredients are added either to reduce the number of
free parameters or to reduce their relevant range. In the first case the predictability of the model
is increased, while in the second case the model becomes more natural. Indeed it would be desirable
to account for the hierarchies of the charged fermion mass ratios and of the quark mixing angles in
terms of an irreducible set of order-one parameters.

In ref. \cite{Kitano:2003cn} Kitano and Li accomplished this goal in a SUSY SO(10) model formulated
in five flat space-time dimensions. The fifth dimension is an interval whose length is of the order
of the inverse GUT scale. Fermions are hosted in three {\bf 16} multiplets living in the full
five-dimensional space-time  while Yukawa interactions, described by matrices with order-one
elements, are localized at one of the branes. The Yukawa couplings for the fermions of the SM
are obtained by convoluting these order-one matrices with the profiles of the fermionic zero-modes.
The resulting picture is essentially equivalent to that produced by many models of fermion masses
such as those based on Froggatt-Nielsen (FN) U(1)$_{\rm FN}$ flavor symmetries
\cite{Froggatt:1978nt} or those relying on the mechanism of partial compositeness
\cite{Kaplan:1991dc,*Csaki:2008zd}. In an SO(10) context one would expect fermions in the same {\bf
16} representation to share the same zero-mode profile, which in the FN language would correspond
to all members of a {\bf 16} having the same FN charge. Such a picture is clearly unrealistic since
it leads to mass ratios for up and down type quarks of the same order of magnitude. The key point of
the Kitano and Li model is that the breaking of SO(10) down to SU(5)$\times$U(1)$_{\rm X}$
determines different profiles for the zero-modes of the different SU(5) components inside the {\bf
16} multiplet. The model becomes flexible enough to account for the different hierarchies observed
in the different charge sectors. Unfortunately the total number of parameters is still very large
since a single matrix of Yukawa interactions localized at one brane is insufficient to correctly
describe the quark mixing and more than one type of Yukawa interactions are required.

In the present paper we improve the model of ref. \cite{Kitano:2003cn} in several respects. The
different type of Yukawa interactions were generated in  \cite{Kitano:2003cn}
through non-renormalizable operators of different dimensionality, which were assumed to give
contributions of the same order. This forces the cut-off scale suppressing higher-dimensional
operators to be close to the GUT scale, thus questioning the domain of validity of the effective
theory. In the version of the model presented here the required Yukawa matrices arise from operators
of the same dimension and the cut-off of the theory can be in principle as large as the Planck mass, making the
contributions from higher dimensional operators negligible, and reducing the theoretical error of
our predictions to the level of the experimental accuracy for most of the data. Moreover we
explicitly address the doublet-triplet splitting problem by choosing a particle content in the Higgs
sector to which the missing partner mechanism
\cite{Masiero:1982fe,*Grinstein:1982um,Babu:2006nf,Babu:2011tw} is directly applicable. Finally, to
check the viability of the model, we perform a fit to an idealized set of data, obtained by running
the observed masses and mixing angles up to the GUT scale. We find good agreement with the data, for large values of $\tan\beta$.
Neutrino masses and mixing angles are reproduced within a type I seesaw mechanism. Normal ordering
of neutrino mass is predicted with the lightest neutrino mass smaller than 5 meV.  All the values
of the Dirac CP phase are found equally preferred and the effective mass of neutrinoless double
beta decay is predicted in the range 0.1-5 meV. It is particularly interesting that the present
SO(10) GUT can give rise to fermion mass matrices similar to the ones obtained in
SU(5)$\times$U(1)$_{\rm FN}$ models \cite{Altarelli:2002sg,*Altarelli:2012ia,Sato:1997hv,*Buchmuller:1998zf}, very effective in reconciling the nearly anarchical pattern of neutrinos with the hierarchical one of
charged fermions. Here we do not aim at a fully realistic model and we deliberately leave apart
several important issues, such as gauge coupling unification and the problem of proton decay.

In the following section, we briefly review the basic framework of Kitano-Li model and explain a mechanism  responsible for creating hierarchies among the different fermions.  In section \ref{model}, we provide a modified version of this model and discuss its essential features in details. The fermion mass relations predicted by the model are discussed and their viability is investigated through detailed numerical analysis in section \ref{numerical}. We finally conclude in section \ref{conclusions}.

\section{Flavour hierarchy from extra-dimension and SO(10) GUT}
\label{flavor-hierarchy-so10}
Let us first review the basic framework of the 5-dimensional (5D) SUSY theory which the 
Kitano-Li model \cite{Kitano:2003cn} is based on. Consider a 5D $N=1$ SUSY U(1) gauge theory
compactified on half a circle $S^1/Z_2$ \cite{Pomarol:1998sd}. It is shown in
\cite{ArkaniHamed:2001tb} that such a theory can conveniently be written in terms of 4-dimensional
superspace formalism. When decomposed into 4D, the 5D vector supermultiplet contains an $N=1$
chiral multiplet $\Phi$ and a vector multiplet $V$. In a similar way, the 5D hypermultiplet consists
of a pair of $N=1$ 4D chiral multiplets $H$ and $H^c$. The U(1) gauge invariant action of
interacting hypermultiplet and vector multiplet is written as
\cite{Pomarol:1998sd,ArkaniHamed:2001tb,Barbieri:2002ic}
\beqa \label{U1-action}
S_5 &=& \int dy~ d^4x~ \left[ \int d^4\theta \left(\partial_y V -
\frac{1}{\sqrt{2}}(\Phi+\bar
{\Phi})\right)^2+ \frac{1}{4} \int (d^2\theta~W^\alpha W_\alpha + {\rm h.c.}) \right. \nonumber
\\
&+& \int d^4\theta~\left(\bar{H} e^{2g_5QV}H + \bar{H}^c e^{-2g_5QV}H^c \right)\nonumber \\
&+&\left. \left( \int d^2\theta~H^c\left({\hat m}+\partial_y -\sqrt{2}g_5Q\Phi\right)H  +{\rm
h.c.}\right)\right], 
\eeqa
where $W^\alpha$ is a field strength, $g_5$ is the 5D gauge coupling constant, ${\hat m}$ is the
bulk mass and $Q$ is the U(1) charge of the chiral multiplet $H$. The mass dimensions are:
$[\Phi]=[H]=[H^c]=+3/2$, $[\hat m]=+1$, $[V]=+1/2$ and $[g_5]=-1/2$.
The vector multiplet $V$ and chiral multiplet $H$ transform as even fields under the $Z_2$ symmetry
while the fields $\Phi$ and $H^c$ are odd. For consistency, the bulk mass parameter ${\hat m}$ is
odd under $Z_2$ and the simplest choice is ${\hat m}(y)=m ~{\rm
sgn}(y)$, $m$ being a real constant. The only interactions of the model allowed by $N=1$ 5D SUSY are
gauge interactions. Indeed $\Phi$ is related to the fifth component of the 5D gauge multiplet and
its interaction with $H$ and $H^c$ is controlled by the gauge coupling constant $g_5$.

The compactification on $S^1/Z_2$ breaks 4D $N=2$ SUSY down to $N=1$ SUSY, thus allowing for a
chiral fermion content. Beyond the bulk action $S_5$ of Eq. (\ref{U1-action})
there can be contributions strictly localized on the branes, which should only respect
$N=1$ SUSY. Here we discuss the theory in the ideal limit of exact $N=1$ SUSY and we neglect soft
SUSY breaking contributions with a characteristic scale in the range $1\div 10$ TeV. 
One can perform the Kaluza-Klein (KK) expansion of 5D
bulk fields and obtain the massless spectrum of the 4D theory using the equations of motion and 
boundary conditions imposed by the $Z_2$  symmetry on different fields.

For the chiral superfield $H(x,y)=\sum_n H_n(x)f_n(y)$, one finds a localized zero-mode profile 
\be \label{zero-mode}
f_0(y)= \sqrt{\frac{2 m}{1-e^{-2 m \pi R}}}e^{-m y}~,
\ee
where $R$ is the compactification radius of the extra dimension. For $m<0$ ($m>0$) the zero-mode
profile $f_0(y)$ of $H$ is localized at the $y=\pi R$ ($y= 0$) brane. This feature can be exploited
to suppress (enhance) the strength of the interactions between such zero mode and 
fields from a Higgs sector localized at the $y=0$ brane. In this way the hierarchical pattern
observed in fermion masses and mixing angles can be explained without appealing to small ad hoc
parameters \cite{ArkaniHamed:1999dc,*Mirabelli:1999ks,Kaplan:2001ga}. The chiral superfield $H^c$
is odd under $Z_2$ and has no zero modes. The massive KK modes have masses $m_n^2=m^2+(n/R)^2$,
above the compactification scale $M_{KK}=1/R$. The vector supermultiplet $V$ has a zero mode
constant in $y$ and given by $1/\sqrt{\pi R}$. Thus the gauge coupling constant $g_4$ of the 4D
effective theory is related to $g_5$ by
\be
g_4=\frac{g_5}{\sqrt{\pi R}}~.
\ee
The chiral multiplet $\Phi$ has no zero mode, but its scalar component can acquire a vacuum
expectation value (VEV).

The above framework is used in \cite{Kitano:2003cn} to construct a grand unified model based on the
SO(10) gauge group. In this model the $N=1$ chiral multiplets $H$ and $H^c$ are replaced
by three copies of ${\bf 16}$ and ${\bf 16}^c$, transforming as ${\bf 16}$ and ${\bf
\overline{16}}$ under SO(10) respectively. The vector supermultiplet, comprising ${\bf 45}_V$ and
${\bf 45}_\Phi$,  transforms in the adjoint of SO(10). 
The breaking of SO(10) down to the SM gauge group is realized in several steps. The VEV of the ${\bf
45}_\Phi$, aligned along the direction of a U(1)$_{\rm X}$ subgroup, breaks SO(10) down to
SU(5)$\times$ U(1)$_{\rm X}$. Since the ${\bf 45}_\Phi$ field is odd under $Z_2$, its VEV has a
non-trivial profile in the fifth dimension, $\langle {\bf 45}_\Phi\rangle = \upsilon_\Phi^{3/2}
sgn(y)$, and generates a D-term for U(1)$_{\rm X}$
\cite{ArkaniHamed:2001tb,Barbieri:2002ic,Kaplan:2001ga}:
\be
-D=\partial_5\langle {\bf 45}_\Phi\rangle=
2\upsilon_\Phi^{3/2}[\delta(y)-\delta(y-\pi R)]~.
\label{Dterm}
\ee
To preserve $N=1$ SUSY such a D-term can be canceled by introducing on the branes new fields charged
under U(1)$_{\rm X}$ \cite{ArkaniHamed:2001tb,Barbieri:2002ic}. In the Kitano-Li model
\cite{Kitano:2003cn} the brane $y=0$ hosts a pair  $({\bf 16}_H,\overline{{\bf 16}}_H)$ of chiral
superfields while another pair $({\bf 16'}_H,\overline{{\bf 16}}'_H)$ is introduced at the brane
$y=\pi R$. Their VEVs are adjusted to exactly  cancel the D-term of Eq. (\ref{Dterm}). In this way,
the U(1)$_{\rm X}$ subgroup is broken near the GUT scale. For this reason the X generator should be
orthogonal to the SM ones and this condition uniquely determines U(1)$_{\rm X}$ inside SO(10). The
other fields needed on the brane $y=0$ are a chiral multiplet ${\bf 45}_H$, which breaks the
residual SU(5) symmetry to the SM gauge group, and ${\bf 10}_H$, which contains a pair of Higgs
doublets. The 5D superpotential of the model is \cite{Kitano:2003cn}:
\beqa\label{W_KL}
{\cal W}_{\rm KL} &=& {\bf 16}^c_i \left[ {\hat m}_i+\partial_y - \sqrt{2} g_5 \,{\bf 45}_{\Phi}
\right]{\bf
16}_i \nonumber \\
& + &\frac{\delta(y)}{\Lambda}~\left[
Y_{ij} {\bf 16}_i {\bf 16}_j{\bf 10}_H  + \frac{(Y_R)_{ij}}{\Lambda} ({\bf 16}_i \overline{\bf
16}_H)({\bf 16}_j 
\overline{\bf 16}_H) +\frac{Y'_{ij}}{\Lambda} {\bf 16}_i {\bf 16}_j {\bf 10}_H {\bf 45}_H   +
...\right] \nonumber \\
& +& \delta(y)~ w_0({\bf 45}_H,{\bf 16}_H, \overline{\bf 16}_H, {\bf 10}_H,...) \nonumber \\
&+&\delta(y-\pi R)~w_{\pi}({\bf 16}'_H, \overline{\bf 16}'_H)~,
\eeqa
where $w_0$ and $w_\pi$ are gauge invariant superpotentials depending only on Higgs supermultiplets
and $\Lambda$ is the cut-off scale of the theory. The basis of ${\bf 16}_i$ is conveniently chosen
so that the bulk mass term of ${\bf 16}_i$ in ${\cal W}_{\rm KL}$ is diagonal. In addition to the
fields contained in the above ${\cal W}_{\rm KL}$, a solution to doublet-triplet splitting problem
through Dimopolous-Wilczeck mechanism \cite{Dimopoulos:1981xm} in the simplest version requires another
${\bf 10}_H$, a pair of ${\bf 16}_H$ and several SO(10) singlet
fields \cite{Barr:1997hq,*Babu:1993we}.

Let us now review the Yukawa sector of the model encoded in the second line of Eq. (\ref{W_KL}).
The first term is responsible for fermion masses of Dirac type. This term would predict an
unrealistic set of masses in the charged fermion sector and for this reason additional contributions
suppressed by more powers of the cut-off scale are needed. One example of such contributions is the
third term in the second line. In ref. \cite{Kitano:2003cn} it is explicitly assumed that all these
contributions effectively give rise to Yukawa matrices in each charge sector, $Y_{u,d,\nu,e}$, that
can be treated as independent. In the second term of the second line, the VEV of ${\bf 16}_H$
generates masses of right-handed neutrinos of the order of the GUT scale, inducing tiny masses for
the light neutrinos through the type I seesaw mechanism.

The Yukawa couplings for the charged fermion zero modes are obtained by convoluting $Y_{u,d,e}$
with the zero-mode profiles, which in turn are controlled by both the bulk masses $m_i$ and the VEV
of ${\bf 45}_\Phi$. Such a VEV generates different contributions to the bulk masses of the SU(5)
components of each ${\bf 16}$ bulk multiplet, proportional to their U(1)$_{\rm X}$ charges. Under
SU(5)$\times$ U(1)$_{\rm X}$ the ${\bf 16}$ decomposes as
\be 
\label{16-decom}
{\bf 16} = 10_{-1} + \bar{5}_{3} +1_{-5}~,
\ee 
where the numbers in subscript represents U(1)$_{\rm X}$ charge of a given SU(5) multiplet: $Q^r_X$. 
Each SU(5) multiplet gets an effective bulk mass $m_i^r$ $(r=10,{\bar 5},1)$ given by
\be
m_i^r=m_i - \sqrt{2}g_5 Q_X^r \upsilon_\Phi^{3/2}~,
\label{y1}
\ee 
that can be expressed in units of the cut-off scale as
\be
m_i^r=\Lambda~ a_i^r
\label{y2}
\ee
in terms of dimensionless quantities
\be
a_i^r=\mu_i-Q_X^r k_X~,~~~~~~~~~\mu_i=\frac{m_i}{\Lambda}~,~~~~~~~k_X=\sqrt{2}\frac{g_5\upsilon_\Phi^{3/2}}{\Lambda}~.
\label{y3}
\ee
The Yukawa  couplings ${\cal Y}_f$ $(f=u,d,e)$ of the charged fermion zero modes are  
\be
{\cal Y}_u=F_{10} Y_u F_{10}~,~~~~~~~{\cal Y}_d=F_{10} Y_d F_{\bar 5}~,~~~~~~~{\cal Y}_e=F_{\bar
5} Y_e F_{10}
\label{y4}
\ee
where the entries of diagonal matrices $F_r$ are the zero-mode profiles evaluated at the $y=0$
brane:
\be
F_r={\tt diag}(n_1^r,n_2^r,n_3^r)~,~~~~~~~~~n_i^r=\sqrt{\frac{2 a_i^r}{1-e^{-2 a_i^r
c}}}~,~~~~~~c=\Lambda \pi R~.
\label{y5}
\ee
The mass matrix of light neutrinos is obtained through the type I seesaw mechanism and is proportional to
\be 
\label{KL-neutrino}
m_\nu \propto F_{\bar 5}~Y_\nu Y_R^{-1} Y_\nu^T~F_{\bar 5}~.
\ee
It was shown in \cite{Kitano:2003cn} that a suitable choice of the VEV of ${\bf 45}_\Phi$ can
generate the
following hierarchy in the profiles:
\be 
\label{analytical-profile}
F_{10}\backsimeq {\tt diag}(\lambda^4,~\lambda^2,~1),~~F_{\bar 5} \backsimeq {\tt
diag}(\lambda,~1,~1) 
\ee
for $10$ and $\bar{5}$ fermions with $\lambda\sim 0.23$. These profiles give rise to realistic
hierarchies in fermions masses and mixing angles, including the neutrinos, even if all the Yukawa
matrices in Eq. (\ref{W_KL}) have anarchical ${\cal O}(1)$ elements. The strong
hierarchy in the profiles of $F_{10}$ compared to $F_{\bar 5}$ provides a qualitative understanding
of the extremely hierarchical spectrum of up-type quarks and the less hierarchical down-type
quarks and charged leptons. The milder hierarchy in the neutrino masses and emergence of the large
mixing angles can also be understood in this way.

\section{A modified Kitano-Li model}
\label{model}
The above framework looks consistent at the qualitative level, but it has not been analyzed on the
quantitative grounds to check its viability and its predictability. In this paper we would like to
address such a question. Moreover, in its present version, the model can be only applied to an
energy range ending very close to the GUT scale, and the effective description it provides could
suffer from large uncertainties coming from the unknown UV completion. Indeed the VEV of ${\bf
45}_H$ breaks SU(5) into the SM gauge group and can be identified with the GUT scale $M_{\rm
GUT}\approx 2 \times 10^{16}$ GeV. The higher-order terms in the second line of Eq. (\ref{W_KL}), as
the one proportional to $Y'$, are suppressed by powers of $M_{\rm GUT}/\Lambda$ and are very small
if $\Lambda \gg M_{\rm GUT}$. In this limit the charged fermion Yukawa interactions on the brane
are dominated by the first term, leading to: $Y_u=Y_d=Y_e= Y$. The down-type quarks and charged
leptons become exactly degenerate in this limit since the zero-mode profiles cannot distinguish
between SM sub-multiplets within $10$ and $\bar{5}$. On the other hand the simple GUT scale
extrapolation of the currently observed values of the masses of down-type quarks and charged leptons
requires $m_b/m_\tau \approx 0.7$, $m_s/m_\mu \approx 0.2$ and $m_d/m_e\approx 2.5$ for almost any
value of $\tan\beta$ \cite{Ross:2007az}. Such large corrections, particularly in the first two
generations, cannot be induced through the higher-dimensional operators unless
$M_{\rm GUT}\sim\Lambda$ is considered and if all the Yukawa couplings in the theory are taken to
be ${\cal O}(1)$ parameters. Taking the cut-off scale $\Lambda$ very close to the $M_{\rm GUT}$
questions the validity of the effective field theory approach which underlies the whole construction
of the model.

The effective theory description can be restored by assuming $\Lambda \gg M_{\rm GUT}$. The
correction in the down-type quarks and charged lepton masses then requires leading-order
contribution in Yukawa interactions which can be achieved either by $\overline{\bf 126}_H$ or ${\bf
120}_H$ or by both. Unlike ${\bf 10}_H$ and $\overline{\bf 126}_H$, the Yukawa interactions of ${\bf
120}_H$ with ${\bf 16}_i$ are anti-symmetric in generation space and hence they introduce less
number of free parameters compared to $\overline{\bf 126}_H$. Keeping this aspect in mind, here we
propose a variant of the Kitano-Li model based on ${\bf 10}_H+{\bf 120}_H$ fields on the brane,
which can account for all the charged fermion masses and mixing angles, as we show through a
detailed quantitative analysis in the next section. The ${\bf 16}_H$ and $\overline{\bf 16}_H$ on
the brane are replaced by ${\bf 126}_H$ and $\overline{\bf 126}_H$, which pick up a VEV at the GUT
scale, solve the D-term problem and generate the masses for right handed (RH) neutrinos through a
leading order term in the  Yukawa interaction. The ${\bf 126}_H$ and $\overline{\bf 126}_H$ also play a crucial
role in solving the doublet-triplet splitting problem through the missing partner mechanism as
described in \cite{Babu:2006nf,Babu:2011tw}. We provide a detailed discussions of the model in this
section.

We use the same field configuration in the bulk as previously used in \cite{Kitano:2003cn} and only
modify the brane sector considerably. We assume as superpotential of the model 
\beqa \label{W}
{\cal W} &=& {\bf 16}^c_i \left[ \hat{m}_i+\partial_y - \sqrt{2} g_5 \,{\bf 45}_{\Phi} \right]{\bf
16}_i \nonumber \\
&+& \frac{\delta(y)}{\Lambda} \left[ Y_{10}^{ij} {\bf 16}_i {\bf 16}_j {\bf 10}_H  + Y_{120}^{ij}
{\bf 16}_i {\bf 16}_j {\bf 120}_H +
Y_{126}^{ij} {\bf 16}_i {\bf 16}_j\overline{\bf 126}_H +...\right] \nonumber \\ 
&+& \delta(y) w_0({\bf 45}_H, {\bf 10}_H, {\bf 126}_H, \overline{\bf 126}_H, {\bf 120}_H)  \nonumber
\\ 
&+& \delta(y-\pi R)w_{\pi}({\bf 126}'_H, \overline{\bf 126}'_H)~. \eeqa
As already discussed, the VEV of ${\bf 45}_\Phi$ breaks the SO(10) symmetry down to
SU(5)$\times$ U(1)$_{\rm X}$ and splits the profiles of the SU(5) sub-multiplets of ${\bf 16}_i$. 
We now discuss in detail the roles played by each of the brane fields in this model.
\begin{itemize}
\item
Under SU(5)$\times$ U(1)$_{\rm X}$ the multiplets ${\bf 10}_H$ and ${\bf 120}_H$ decompose as:
\beqa 
\label{10-120-decomp}
{\bf 10}_H &=& 5_2+\overline{5}_{-2}~, \nonumber \\
{\bf 120}_H &=& 5_2+\overline{5}_{-2}+10_{-6}+\overline{10}_6+ 45_2+\overline{45}_{-2}~.
\eeqa
The ${\bf 10}_H$ contains a pair of weak doublets, one in $5$ and the other in $\bar5$, which
transforms as a pair of Higgs doublets in the Minimal Supersymmetric Standard Model (MSSM), while
${\bf 120}_H$ contains two pairs of such doublets, one pair of doublets residing
in $45$ and $\overline{45}$ of SU(5). We assume that these doublets get mixed with each other
through the couplings in the superpotential $w_0$ and that only one linear combination of them
remains light and plays the role of MSSM Higgs doublets. A natural solution of the doublet-triplet
splitting problem leading to such light pair of doublets is offered by the missing partner
mechanism in this model, as we discuss it later in detail. Since the light doublets are admixtures of
doublets in ${\bf 10}_H$ and ${\bf 120}_H$, the Yukawa couplings of charged SM fermions are linear
combinations of $Y_{10}$ and $Y_{120}$. Such relations were derived explicitly in
\cite{Barbieri:1979ag} and we write them in the next section. It is well known that a pair of MSSM
doublets residing in $45$ and $\overline{45}$ of SU(5) distinguishes the Yukawa couplings of
down-type quarks from those of the charged leptons. 
\item 
The $\overline{\bf 126}_H$ representation of SO(10) decomposes under SU(5)$\times$ U(1)$_{\rm X}$
as
\be 
\overline{\bf 126}_H=1_{10}+5_2+\overline{10}_6+15_{-6}+\overline{45}_{-2} +50_2~.
\ee
An analogous decomposition for the ${\bf 126}_H$ holds. The pair $(\overline{\bf 126}_H,{\bf
126}_H)$  replaces the pair $({\bf 16}_H,\overline{\bf 16}_H)$ used by Kitano and Li and plays a
similar role. The VEVs of the SU(5) singlets residing in ${\bf 126}_H$, $\overline{\bf 126}_H$ are
used to cancel the D-term on the branes that arises from the VEV of ${\bf 45}_\Phi$. The vanishing 
of the D-term requires \cite{ArkaniHamed:2001tb,Barbieri:2002ic}
\beqa 
\label{Dterm-cond}
0=-D_{\rm U(1)_X}&=& \delta(y) \left[2\upsilon_\Phi^{3/2} + 10 g_5 (|\langle {\bf 126}_H
\rangle|^2-|\langle \overline{\bf 126}_H \rangle|^2)\right] \nonumber \\
&-&\delta(y-\pi R) \left[2\upsilon_\Phi^{3/2} - 10 g_5 (|\langle {\bf 126}'_H \rangle|^2-|\langle
\overline{\bf 126}'_H \rangle|^2)\right]~,
\eeqa
where we identify the gauge coupling constant of U(1)$_{\rm X}$ with $g_5$. Clearly, this breaks the
U(1)$_{\rm X}$ symmetry and reduces the rank of the residual gauge symmetry. The VEV of
$\overline{\bf 126}_H$ also generates the masses for the RH neutrinos through the Yukawa interaction
term proportional to $Y_{126}$ of Eq. (\ref{W}). Note that $\overline{\bf 126}_H$ also contains a
pair of weak doublets. However such a pair is assumed to be as heavy as the other submultiplets of 
$\overline{\bf 126}_H$ as required by the missing partner mechanism for solving the doublet-triplet
splitting problem, as we discuss below. In this way, $\overline{\bf 126}_H$ does not contribute to
the charged fermion masses.
\item
The decomposition of ${\bf 45}_H$ is given by
\be
{\bf 45}_H=1_0+10_4+\overline{10}_{-4}+24_0~.
\ee
Note that ${\bf 45}_H$ contains an adjoint of SU(5) and can trigger the SU(5) breaking down to the
SM gauge symmetry. This cannot be achieved by ${\bf 45}_\Phi$ in the bulk because the VEV of the
24-plet of SU(5) residing in ${\bf 45}_\Phi$ would induce a non-vanishing D-term corresponding to
U(1)$_{\rm Y}$. Such a D-term cannot be canceled without the breaking of U(1)$_{\rm Y}$ and hence we
need a ${\bf 45}_H$ to break SU(5).
\end{itemize}

The above Higgs content on the brane naturally solves the doublet-triplet splitting problem through
the missing partner mechanism as pointed out in \cite{Babu:2011tw}. In this mechanism, a set of
``light'' fields is considered, with an assumption that they get masses only through interactions
with ``heavy'' fields. In other words, the mechanism assumes the absence of bare mass terms for the
light Higgs sector. In the above model, ${\bf 10}_H$ and ${\bf 120}_H$ fields can be considered as
light, while ${\bf 126}_H$, $\overline{\bf 126}_H$ and ${\bf 45}_H$ are considered as the heavy
ones. As can be seen from the decomposition under the SM gauge group, the light fields contain
three pairs of weak doublets and three pairs of color triplets. The heavy fields contain the same
number of triplets but only two pairs of doublets.

The unequal content of doublets and triplets in the heavy sector arises from the $50$,
$\overline{50}$ of SU(5) residing in ${\bf 126}_H$, $\overline{\bf 126}_H$ which contain only
triplets. One assumes that there is no GUT scale bare mass terms for the light fields so that
different sub-multiplets of the light fields get masses through their interactions with ${\bf
126}_H$, $\overline{\bf 126}_H$ and ${\bf 45}_H$. Such interactions can arise in $w_0$, for example
\be 
\label{DT-potential}
w_0 =  {\bf 120}_H~{\bf 126}_H~{\bf 45}_H +{\bf 120}_H~\overline{\bf 126}_H~{\bf 45}_H +
\frac{1}{\Lambda} {\bf 10}_H~{\bf 126}_H~{\bf 45}_H^2 +
\frac{1}{\Lambda} {\bf 10}_H~\overline{\bf 126}_H~{\bf 45}_H^2 +... 
\ee
The doublets and triplets from the different heavy and light fields get mixed with each other when
${\bf 45}_H$ takes a VEV. The three triplets from the light fields get mixed with the same number of
triplets in the heavy fields and all of them obtain GUT scale masses. On the other hand, one
combination of weak doublets in the light sector remains massless since the heavy sector contains
only two of such doublets. It is also shown in \cite{Babu:2011tw} that the other sub-multiplets in
${\bf 120}_H$ also get mixed with their counterparts in the heavy fields and all of them become
massive. Hence one finds only one linear combination of weak doublets from the ${\bf 10}_H+{\bf
120}_H$ which remains light and can be used as the MSSM Higgs doublets.

The above scalar content, {\it i.e.} ${\bf 10}_H+{\bf 120}_H$ fields as the light fields and
${\bf 45}_H+{\bf 126}_H+\overline{\bf 126}_H$ as the heavy fields, is the most economical among the
other possibilities \cite{Babu:2011tw} of light and heavy fields which provide a solution to
the doublet-triplet splitting problem through the missing partner mechanism in SO(10). However, in 4D
SO(10) theories, ${\bf 10}_H$ and ${\bf 120}_H$ alone do not lead to realistic charged fermion
masses and quark mixing angles as first pointed out in \cite{Lavoura:2006dv} through a numerical
study. The limited numbers of Yukawa couplings were found unable to reproduce appropriate
hierarchies in the charged fermion masses. This is not the case in the present model
as we show it later explicitly through a detailed numerical analysis. The zero-mode profiles of the
different fermions generated from the compactification of an extra dimension in this model relax the
tension that exists in pure 4D theories. Before we proceed to a quantitative analysis of the fermion mass spectrum in the above
framework, we discuss the range of validity of the effective field theory approach on which this model is based.
\begin{figure}[t]
\centering
\subfigure{\includegraphics[width=0.5\textwidth]{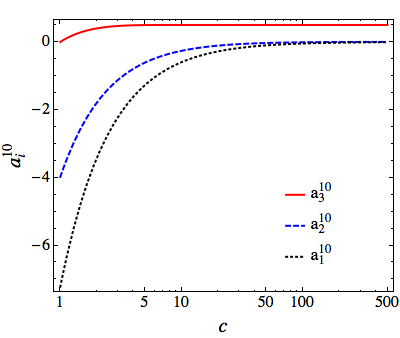}}
\caption{The bulk mass parameters $a_i^{10}$ as functions of $c=\pi \Lambda/M_{KK}$ as required from
the hierarchy in up-type quark masses. The dotted, dashed and solid lines correspond to $a_1^{10}$,
$a_2^{10}$ and $a_3^{10}$ respectively.}
\label{fig1}
\end{figure}

The effective Yukawa couplings in 4D and the light neutrino mass matrix are as in the Kitano-Li
original model, Eqs. (\ref{y1}--\ref{KL-neutrino}), where now $Y_R=Y_{126}$ and $Y_{u,d,\nu,e}$ are
linear combinations of $Y_{10}$ and $Y_{120}$. All the Yukawa couplings $(Y_{10})_{ij}$,
$(Y_{120})_{ij}$ and $(Y_{126})_{ij}$ are assumed to be of order one. The behaviour of the fermion
zero-modes at $y=0$ brane can be classified according to the values of the bulk mass parameters and
$c$:
\begin{itemize}
\item[] ~~~~~~~~~~~~~~~~~ for $a^r_i>0$ and $|a^r_i| c\gtrsim 1$,~~~~~~$n^r_i \approx
\displaystyle\sqrt{2 a^r_i}$ 
\item[] ~~~~~~~~~~~~~~~~~ for $a^r_i<0$ and $|a^r_i| c\gtrsim 1$,~~~~~~$n^r_i \approx
\displaystyle\sqrt{2 |a^r_i|}~e^{\displaystyle-|a^r_i| c}$ 
\item[] ~~~~~~~~~~~~~~~~~ for $a^r_i\lessgtr 0$ and $|a^r_i| c  < 1$,~~~~~~$n^r_i \approx
\displaystyle\frac{1}{\sqrt{c}} $~.
\end{itemize}
The parameter $c=\pi\Lambda/M_{KK}$ represents the cut-off scale in units of the KK scale
and, to consistently describe the first few KK modes within our effective theory, we take $c\ge
10$. To neglect higher-dimensional operators contributing to fermion masses we will show that
a value of $c$ larger than 10 is required. To reproduce the large top Yukawa coupling we have to
take $a^{10}_3=\mu_3+k_X\approx y_t/2$. The hierarchy among the first, second and third
generations of quarks can be reproduced by choosing $\mu_3\approx y_t/2$ and $|k_X|,|\mu_{1,2}|\ll 1$.
For example, the values of $a^{10}_i$ required to
generate $(n^{10}_1,n^{10}_2,n^{10}_3)=(\lambda^4,\lambda^2,1)$ are shown in Fig. \ref{fig1}, as a
function of $c$. For large $c$, $a^{10}_3$ approaches to $0.5$ while $|a^{10}_{1,2}|$ go like $1/c$. In terms of our input parameters, we approximately have $\mu_3\approx 0.5$, while
$|k_X|,~|\mu_{1,2}|$ are ${\cal O}(1/c)$.

The parameter $c$, describing the gap between the cut-off scale and the KK scale, characterizes the
domain of validity of our effective theory. Here we estimate how large $c$ can be in our model and
how small can be the ratio $M_{\rm GUT}/\Lambda$, which controls the non-leading contributions to
the Yukawa interactions on the $y=0$ brane.  A relation between $c$ and the GUT scale parameters such as  $k_X$ can be derived from the phenomenological requirement $|\mu_{1,2}| \sim |k_X| \approx 1/c$, needed to successfully fit the fermion spectrum. There are several scales relevant to the breaking of the grand unified symmetry SO(10): $\upsilon_\Phi$, $\langle{\bf 126}_H \rangle$, $\langle\overline{\bf 126}_H \rangle$, $\langle{\bf 45}_H \rangle$. In first approximation we make no distinction among them and we assume $M_{\rm
GUT}\approx \upsilon_\Phi$. From Eq. (\ref{y3}) and using $|k_X|\sim 1/c$, one can express the VEV of the ${\bf 45}_\Phi$ in terms of  $c$ as: 
\be
\upsilon_\Phi=\left(\frac{1}{2g_4^2}\right)^{1/3}\frac{\Lambda}{c}~,
\label{zeta}
\ee
where we use $g_5^2=\pi R g_4^2$. Considering a dimensionless coupling $g_4 \sim {\cal O}(1)$, one finds \be
\frac{M_{\rm GUT}}{\Lambda}\approx\frac{1}{c}~,
\ee
showing that in the preferred region of parameter space the GUT scale and the KK scale are close to
each other.  To conveniently suppress the higher order contribution to the Yukawa interactions on the
$y=0$ brane we can take $c=100$ and the cut-off $\Lambda$ approximately corresponds to the Planck
scale. The bulk mass parameters relevant to our analysis, $m_i$ and $\upsilon_\Phi$, are all around the GUT
scale, except $m_3$ which should be relatively close to $\Lambda$.

\section{Fermion mass relations and Numerical Analysis}
\label{numerical}
We now derive the effective mass matrices of fermion zero modes in the model and discuss their
viability through a detailed numerical analysis. As noted earlier, the light fields ${\bf 10}_H$ and
${\bf 120}_H$ respectively contain one and two pairs of MSSM-like Higgs doublets. As it is arranged
by the missing partner mechanism, one pair of their linear combinations remains massless and plays
the role of the MSSM Higgs doublets, namely $H_u$ and $H_d$. Hence each of the doublets $H_{u,d}^1
\in {\bf 10}_H$ and $H_{u,d}^{2,3} \in {\bf 120}_H$ has a component of $H_u$ or $H_d$ which can
conveniently be parametrized in terms of mixing parameters $\alpha_i$ and $\bar\alpha_i $ such that
$H_u^i = \alpha_i H_u$ and $H_d^i = \bar\alpha_i H_d$. The appropriate normalizations of $H_u$ and
$H_d$ then require
\be  \label{normalization}
\sum_{i=1}^3 |\alpha_i|^2 = \sum_{i=1}^3 |\bar\alpha_i|^2 = 1. \ee 
The VEVs of MSSM Higgs doublets are fixed by the electroweak symmetry breaking scale and are denoted
by $\langle H_u \rangle =\upsilon \sin\beta$ and $\langle H_d \rangle =\upsilon \cos\beta$ with
$\upsilon = 174$ GeV.

The brane Yukawa couplings of Dirac type fermions are obtained as the linear combinations of only
two matrices, $Y_{10}$ and $Y_{120}$, weighted by the appropriate Clebsch-Gordan (CG) coefficients
and Higgs mixing parameters $\alpha_i$, $\bar\alpha_i$
\cite{Barbieri:1979ag,Aulakh:2008sn}:
\beqa \label{Yukawa-sum-rule}
Y_u &=& c_1^u \alpha_1 Y_{10} + (c_2^u \alpha_2 + c_3^u \alpha_3) Y_{120}~, \nonumber \\
Y_d &=& c_1^d \bar\alpha_1 Y_{10} + (c_2^d \bar\alpha_2 + c_3^d \bar\alpha_3) Y_{120}~, \nonumber \\
Y_\nu  &=& c_1^\nu \alpha_1 Y_{10} + (c_2^\nu \alpha_2 + c_3^\nu \alpha_3) Y_{120}~, \nonumber \\
Y_e &=& c_1^e \bar\alpha_1 Y_{10} + (c_2^e \bar\alpha_2 + c_3^e \bar\alpha_3) Y_{120}~, \eeqa
where $Y_{u,d,e,\nu}$ are Yukawa matrices for up-type quarks, down-type quarks, charged leptons and
Dirac neutrinos. The $Y_{10}$ and $Y_{120}$ are symmetric and anti-symmetric matrices respectively
in generation space. The CG coefficients can be read as $c_1^u=c_1^d=c_1^e=c_1^\nu=2\sqrt{2}$,
$c_2^u=c_2^d=c_2^e=c_2^\nu=-2\sqrt{2}$ and $-3 c_3^u=-3 c_3^d=c_3^e=c_3^\nu=-2i\sqrt{6}$
\cite{Aulakh:2008sn}. A doublet $H_d^3$ residing in ${\bf 120}_H$ couples to the charged leptons and
down-type quarks with different CG coefficients and provides the correction to the wrong mass relation
$Y_d=Y_e$ predicted in the presence of only ${\bf 10}_H$. The above $Y_u$, $Y_d$ and $Y_e$ are
substituted in Eq. (\ref{y4}) to obtain the effective Yukawa matrices ${\cal Y}_{u}$, ${\cal Y}_{d}$
and ${\cal Y}_{e}$ of the charged fermion zero modes at the GUT scale.

The RH neutrinos get mass from the Yukawa interactions of ${\bf 16}$ with ${\bf \overline{126}}_H$
when the SU(5) singlet in ${\bf \overline{126}}_H$ acquires a VEV. The mass matrix of the RH
neutrino zero modes takes the form
\be \label{MR}
M_R \equiv \upsilon_{R}~F_1 Y_{126} F_1~. \ee
Note that $\upsilon_R \equiv \langle {\bf \overline{126}}_H\rangle$ also contributes in canceling
the D-terms and is required to be close to the GUT scale in the absence of any fine tuning. The
mass matrix for the light neutrinos generated by the canonical seesaw mechanism can be written as
\be \label{seesaw}
M_\nu = - \frac{\upsilon^2 \sin^2 \beta}{\upsilon_R}~F_{\bar5}~Y_\nu Y^{-1}_{126}
Y_\nu^T~F_{\bar5} \equiv - \frac{\upsilon^2 \sin^2 \beta}{\upsilon_R}~{\cal Y}_\nu~.\ee
The hierarchy in light neutrino masses is solely governed by the matrix $F_{\bar 5}$ 
and, as we will see,  it leads to relatively less hierarchical neutrinos in comparison to the 
charged fermions as arranged by SO(10) breaking in the bulk. Also, the origin of large (but not
special values of) lepton mixing angles is apparent in this case. The seesaw mechanism is often
seen as the origin of small hierarchies and large mixing in the neutrinos compared to the quark
sector. For example, in the mechanism known as seesaw enhancement \cite{Smirnov:1993af}, such a
difference can be realized if RH neutrinos have strong hierarchy or they are almost degenerate and
Dirac neutrino Yukawas as hierarchical in structure as those of  up-type quarks. We would like to
emphasize here that hierarchy in the RH neutrino masses in this model is not responsible for
enhancement in the leptonic mixing angles as can be seen from Eq. (\ref{seesaw}). In fact since the
RH neutrino unifies with other fermions in SO(10), the hierarchy in their masses can be predicted
from the common bulk mass parameters once the appropriate profiles of charged fermions are
obtained.

In principle the SU(2)$_{\rm L}$ triplet contained in ${\bf \overline{126}}_H$ can generate an additional
contribution to the neutrino masses through  the so-called type II seesaw mechanism
\cite{Lazarides:1980nt,*Schechter:1980gr,*Mohapatra:1980yp}. The coupling of ${\bf
\overline{126}}_H$ to SO(10) multiplets containing light Higgs doublets induces an
effective VEV for such a triplet of order $\upsilon^2\sin^2\beta/M_{\rm GUT}$. In our model, the
missing partner mechanism assumes the absence of a direct coupling among ${\bf \overline{126}}_H$
and the two SO(10) multiplets hosting the light Higgs doublets. For example, ${\bf
\overline{126}}_H$ does not have an SO(10) invariant tree level couplings with two ${\bf 10}_H$ or
with two ${\bf 120}_H$ or with ${\bf 10}_H$ and ${\bf 120}_H$. Further, any such couplings through
higher-dimensional operators are assumed to be absent in the missing partner mechanism. Hence, we do
not expect type II seesaw contribution to the light neutrino masses and consider only type I seesaw
as the mechanism at work for the light neutrino masses in the following analysis.

\subsection{Fitting the fermion mass spectrum: a viability test}
The Yukawa matrices, ${\cal Y}_{u,d,e,\nu}$ which follow from Eqs. (\ref{y4},\ref{Yukawa-sum-rule},\ref{seesaw})  are predicted at the GUT scale and we compare them with a
representative set of data obtained by extrapolating the measured values of fermion masses and
mixing parameters. This strategy has been largely followed in the studies based on varieties of
SO(10) models in four dimensions, see  \cite{Bertolini:2006pe,
*Grimus:2006rk,*Altarelli:2010at,*Joshipura:2011nn,*Dueck:2013gca,*Altarelli:2013aqa} for examples. 
It is clear that an extrapolation over more than 14 orders of magnitude is  potentially affected
by large uncertainties. This is even more true in our model where SUSY breaking effects
have been neglected. We are aware that the data we are going to fit at the GUT scale might not
faithfully represent the low-energy experimental quantities and that the best fit values of the
input parameters we will obtain might considerably change depending on the spectrum of the SUSY
particles and the other heavy modes at the GUT scale. We are more interested in the performances of
the present model, and we are confident that if it can successfully reproduce a representative set
of data, it will also be successful if this set is modified to account for a more realistic
framework.

The renormalization group evolution (RGE) of fermion masses in the MSSM primarily depends on the
SUSY breaking scale $M_{\rm SUSY}$ and $\tan\beta$. In our numerical study we use, as an idealized
set of data at the GUT scale, extrapolated values of the charged fermion masses and the quark mixing
parameters from \cite{Ross:2007az} which, in a 2-loop analysis, assumes an effective SUSY breaking
scale $M_{\rm SUSY}= 0.5$ TeV and considers different values of $\tan\beta$. To assess the dependence
on $\tan\beta$ we will focus on two cases: $\tan\beta=10$ and $\tan\beta=50$.
For the neutrino masses and mixing angles, we use their low-energy values obtained by one of the
recent global fits \cite{Capozzi:2013csa,GonzalezGarcia:2012sz,Forero:2014bxa} ignoring the RGE
effects from the low scale to the GUT scale. The running of the neutrino masses and mixing angles in the
MSSM is known to be negligible in case of $\tan\beta \lsim 30$. It remains small even for large
$\tan\beta$ if the neutrino mass spectrum is hierarchical. As we show later in this section, the
model analyzed here favours large $\tan\beta$ and strongly hierarchical spectrum for neutrinos
with the lightest neutrino mass $\lsim 0.005$ eV at the GUT scale. For such a hierarchical neutrino
spectrum, the RGE running in the neutrino parameters can be considered negligible to a good
approximation \cite{Chankowski:2001mx,Antusch:2003kp,*Antusch:2005gp}. The GUT scale values of
different observables we use in our analysis are listed in Table \ref{GUT-data}. We would like to
note that the given extrapolated values do not take into account the threshold corrections in the
fermion masses and mixing angles arising from the SUSY breaking. Estimation of such corrections
requires precise knowledge of sparticle spectrum which depends on the exact details of SUSY breaking
mechanism \cite{Hall:1993gn,*Blazek:1995nv,*Pierce:1996zz} which is not studied here. Further, the
threshold corrections may also arise at the GUT scale which can be estimated only if the complete mass
spectrum of all the GUT multiplets is known. As it is often done in the similar kind of analysis
\cite{Bertolini:2006pe,*Grimus:2006rk,*Altarelli:2010at,*Joshipura:2011nn,*Dueck:2013gca,
*Altarelli:2013aqa}, we ignore these effects and assume that if the model under consideration can
fit the idealized GUT scale data listed in Table \ref{GUT-data} then it would also be compatible
with the real data.

\begin{table}[t]
\begin{small}
\begin{center}
\begin{tabular}{ccc}
 \hline
 \hline
  Observables ~~~~&~~~~$\tan\beta=10$ ~~~~&~~~~ $\tan\beta=50$~~~~\\
 \hline
$y_t $ ~&~ $  0.48\pm 0.02$ ~&~ $ 0.51\pm 0.03$ \\
$y_b $ ~&~ $  0.051\pm 0.002$ ~&~ $ 0.37\pm 0.02$ \\
$y_{\tau} $ ~&~ $  0.070\pm 0.003$ ~&~ $ 0.51\pm 0.04$ \\
 $m_u/m_c$ ~&~ $  0.0027 \pm 0.0006 $ ~&~ $0.0027\pm 0.0006$ \\
 $m_d/m_s$ ~&~ $0.051 \pm 0.007$ ~&~ $0.051 \pm 0.007$ \\
 $m_e/m_{\mu}$ ~&~ $0.0048 \pm 0.0002$ ~&~ $0.0048 \pm 0.0002$ \\
 $m_c/m_t$ ~&~ $0.0025\pm 0.0002 $ ~&~ $0.0023\pm 0.0002 $ \\
 $m_s/m_b$ ~&~ $0.019 \pm 0.002$ ~&~  $0.016 \pm 0.002$ \\
 $m_\mu/m_\tau$ ~&~ $ 0.059 \pm 0.002$  ~&~ $ 0.050 \pm 0.002$ \\
  \hline
 $|V_{us}|$ & \multicolumn{2}{c}{$0.227 \pm  0.001$}  \\
 $|V_{cb}|$ & \multicolumn{2}{c}{$0.037 \pm 0.001 $}  \\
 $|V_{ub}|$ & \multicolumn{2}{c}{$0.0033 \pm 0.0006 $}  \\
 $J_{CP}$ & \multicolumn{2}{c}{$0.000023 \pm 0.000004 $}  \\
 \hline
 $\Delta_S/10^{-5}$ eV$^2$& \multicolumn{2}{c}{$ 7.54 \pm 0.26$ ~(NO or IO)}  \\
 $\Delta_A/10^{-3}$ eV$^2$ & \multicolumn{2}{c}{$2.44\pm 0.08$~(NO)~~~$2.40\pm 0.07$~(IO)}   \\
 $\sin^2\theta_{12}$ & \multicolumn{2}{c}{$0.308\pm0.017$~(NO or IO)}   \\
 $\sin^2\theta_{23}$ & \multicolumn{2}{c}{$0.425\pm0.029$~(NO)~~~$0.437\pm0.029$~(IO)}  \\
 $\sin^2\theta_{13}$ & \multicolumn{2}{c}{$0.0234\pm0.0022$~(NO)~~~$0.0239\pm0.0021$~(IO)}   \\
\hline
\hline
\end{tabular}
\end{center}
\end{small}
\caption{The GUT scale values of the charged fermion masses and quark mixing parameters from 
\cite{Ross:2007az} that we use in our analysis. The lepton mixing angles and solar and atmospheric
mass differences are taken from a global fit analysis \cite{Capozzi:2013csa} ignoring the running
effects. NO (IO) stands for the normal (inverted) ordering in the neutrino masses.}
\label{GUT-data}
\end{table}

We now proceed to the details of fitting procedure. Let us first calculate the total number of free
parameters in this model. As already mentioned, $Y_{10}$ and $Y_{126}$ are complex symmetric
matrices and $Y_{120}$ is a complex anti-symmetric matrix in generation space. Without loss of
generality, we can absorb three phases from $Y_{10}$ into the ${\bf 16}_i$ by a suitable redefinition
${\bf 16}_i \to e^{i\alpha_i}{\bf 16}_i $. Hence $Y_{10}$ can be parametrized in terms of 9
real parameters, namely 3 real diagonal elements and 3 complex off-diagonal ones. $Y_{126}$ ($Y_{120}$)
contains 12 (6) real parameters. All the Yukawa couplings are regarded as generic order-one
quantities in this model and we constrain them within a narrow range, {\it i.e.} $0.5 \le
|(Y_{10})_{ij}|, ~|(Y_{126})_{ij}|,~|(Y_{120})_{ij}| \le 1.5$ with arbitrary phases. The parameters
$\alpha_1$, $\bar\alpha_1$ in Eq. (\ref{Yukawa-sum-rule}) can be taken real without loss of
generality and can be obtained from $\alpha_{2,3}$, $\bar\alpha_{2,3}$ using the normalization
condition, Eq. (\ref{normalization}). This leaves eight real parameters in $\alpha_i$ and
$\bar\alpha_i$ with the constraints $|\alpha_i|,~|\bar\alpha_i|<1$, a VEV $\upsilon_R$ in Eq.
(\ref{seesaw}) and four real parameters in the profiles of zero-mode fermions as described in Eq.
(\ref{y3}). In total, we have 27 ${\cal O}(1)$ real parameters as  Yukawa couplings and  other 13
real parameters which should correctly reproduce the 18 observables listed in Table \ref{GUT-data}
if the model is viable.

The values of the free parameters of the model are estimated using the $\chi^2$ optimization technique
which is widely used in \cite{Bertolini:2006pe,
*Grimus:2006rk,*Altarelli:2010at,*Joshipura:2011nn,*Dueck:2013gca,*Altarelli:2013aqa} for similar
kind of analysis. We define a $\chi^2$ function
\be \label{chi-square}
\chi^2=\sum_i^{n}\left(\frac{P_{i} (x_1,x_2,..,x_m)- O_i}{\sigma_i}\right)^2, \ee
where $P_i$ are the observable quantities derived from Eqs. (\ref{y4}, \ref{Yukawa-sum-rule},
\ref{seesaw}) as complex nonlinear functions of the free parameters of the model. $O_i$ and
$\sigma_i$ are the GUT scale central values and $1\sigma$ deviations respectively of the
corresponding quantities listed in Table \ref{GUT-data}. The effective Yukawa matrices ${\cal
Y}_{u,d,e,\nu}$ are numerically diagonalized and we obtain the diagonal Yukawas as the eigenvalues of
${\cal Y}_f$ for each sector. For example, the eigenvalues of ${\cal Y}_u$ correspond to 
$y_u,~y_c,~y_t$. The absolute values of the third generation Yukawas and appropriate ratios for the
first two generations are included in the $\chi^2$ to fit them to their extrapolated experimental values. The
quark and lepton mixing parameters are also evaluated in a similar way. For simplicity, we include
a ratio $\Delta_S/\Delta_A$ in $\chi^2$ instead of $\Delta_S$ and $\Delta_A$ individually. As can be
seen from Eq. (\ref{seesaw}), such  a ratio and lepton mixing parameters do not depend on
$\upsilon_R$. The value of $\upsilon_R$ can be obtained later from the fit using the absolute scale
of atmospheric neutrino oscillation. This allows us to remove one observable and one free parameter
from the fit. The $\chi^2$ function contains only dimensionless quantities. It is then numerically
minimized using the downhill-simplex algorithm incorporated in the software tools MINUIT developed
by CERN to determine the best fit values of the parameters $x_i$. From the  fitted parameters, one
can derive the predictions for the various observables which have not been measured yet such as
Dirac CP phase in the leptonic sector, the lightest neutrino mass, the effective mass of
neutrinoless double beta decay.

As a preliminary step we fit all the $39$ free parameters of the model using  $17$ observables
mentioned above.  Even though many of the free parameters are restricted within narrow ranges of
${\cal O}(1)$ or should face additional constraints like the one of Eq. (\ref{normalization}),
the number of free parameters are significantly larger than the number of  observables.
Nevertheless, considering the complex and non-linear dependence of the observable quantities
from the input parameters, it is not completely evident that the model can successfully fit the
data. We have carried out the $\chi^2$ minimization for two different data set corresponding to
$\tan\beta=10$ and $50$. Also, each case is analyzed for different ordering of the neutrino masses,
{\it i.e.} normal (NO) and inverted (IO). The results of fits are reported in Table
\ref{fit-results}. We obtain very good fit to the data in case of $\tan\beta=50$ and for both the NO
and IO in neutrino masses. In particular, the NO case results into a very good fit in which all the
observables from the theory fall well into the experimentally allowed range, as can be seen from
Table \ref{fit-results}. The predictions for various observables obtained at the best fit are also
listed in the table. The set of input parameters at the minimum of $\chi^2$ are collected in
Appendix for both cases. From the best fit in NO case, one obtains a hierarchical profile matrix $F_{10}
\approx {\tt diag.}(\lambda^{3.7}, \lambda^{2.4},1)$ for 10-plets and a relatively less hierarchical
$F_{\bar5} \approx {\tt diag.}(\lambda^{1.5}, \lambda^{0.9},1)$ for $\bar 5$-plets as it was
expected from the SO(10) breaking effects in the bulk.  Such an effect is mostly due to the parameter
$k_X$ which contributes universally in different flavours. As it can be seen from the best fit values
reported in the Appendix, $k_X$ is required to be ${\cal O}(\mu_{1,2})$ in order to distinguish
between the mass hierarchies among the first and second generations of fermions residing in 10 and
${\bar 5}$ of SU(5). As a result, the bulk mass parameter of the third generation $\mu_3$ dominates over $k_X$
and the effective bulk masses $a^r_3$ are nearly equal for both $r=10$ and  $r={\bar 5}$.
Thus an approximate $t-b-\tau$ Yukawa unification at
the GUT scale is enforced. As it is well known, the bottom and tau Yukawas unify with that of top
quark for large $\tan\beta \ge 45$ \cite{Ananthanarayan:1991xp,*Ananthanarayan:1992cd} and hence the model provides a good fit to the data only
for $\tan\beta=50$. We obtain very poor fits for $\tan\beta=10$ corresponding to
$\chi^2_{\rm min} \approx 110$ for NO and $\chi^2_{\rm min} \approx 280$ for IO. The large values of the
$\chi^2_{\rm min}$ in these cases are mainly due to the top, bottom and tau Yukawas, which cannot be
fitted simultaneously to their extrapolated values.

\begin{table}[!ht]
\begin{small}
\begin{center}
\begin{tabular}{c|cc|cc}
 \hline
 \hline
	& \multicolumn{2}{|c|}{Normal ordering} & \multicolumn{2}{|c}{Inverted ordering} \\
~~~\textbf{Observable}~~~ & ~~~\textbf{Fitted value}~~ & ~~\textbf{Pull}~~~&~~~ \textbf{Fitted
value}~~ &
~~\textbf{Pull}~~~\\
 \hline
$y_t $ & 0.51 & 0 & 0.54 & 1.00 \\
 $y_b$ & 0.37 & 0 & 0.37 & 0 \\
 $y_{\tau}$ & 0.51 & 0 & 0.47 & -1.00 \\
$m_u/m_c$ & 0.0027 & 0 & 0.0031 & 0.67\\
$m_d/m_s$& 0.051 & 0 & 0.045 & -0.86\\
$m_e/m_\mu$ & 0.0048 & 0 & 0.0048 & 0 \\
$m_c/m_t$ & 0.0023 & 0 & 0.0023 & 0 \\
$m_s/m_b$& 0.016 & 0 & 0.015 & -0.50 \\
$m_\mu/m_\tau$& 0.050 & 0  & 0.049 & -0.50 \\
\hline
$|V_{us}|$& 0.227 & 0 & 0.227 & 0\\
$|V_{cb}|$& 0.037 & 0 & 0.038 & 1.00 \\
$|V_{ub}|$& 0.0033 & 0  &  0.0030 & -0.50 \\
$J_{CP}$ & 0.000023 & 0 & 0.000021 & -0.51 \\ 
\hline
 $\Delta_S/\Delta_A$  &  0.0309 & 0 & 0.0320 & 0.73\\
$\sin^2 \theta _{12}$ & 0.308 & 0  &  0.309 & 0.06 \\
$\sin^2 \theta _{23}$ & 0.425 & 0 & 0.435 & -0.07  \\
$\sin^2 \theta _{13}$ & 0.0234 & 0 &  0.0237 & -0.10 \\
\hline
\hline
$\chi^2_{\rm min}$ & & $\approx 0$& &  $\approx 5.75$\\
\hline
\hline
           & \multicolumn{2}{|c|}{\textbf{Predicted value}} &\multicolumn{2}{|c}{\textbf{Predicted value}}\\
$m_{\nu_{\rm lightest}}$ [meV] & \multicolumn{2}{|c|}{$0.08$} & \multicolumn{2}{|c}{$2.15$} \\
$|m_{\beta\beta}|$ [meV]& \multicolumn{2}{|c|}{1.63} &\multicolumn{2}{|c}{$30.4$}\\
$\sin \delta^l_{CP}$ &\multicolumn{2}{|c|}{0.265} &\multicolumn{2}{|c}{0.510}\\
$M_{N_1}$ [GeV]& \multicolumn{2}{|c|}{$3.85 \times10^6$} & \multicolumn{2}{|c}{$1.13 \times10^4$}\\
$M_{N_2}$ [GeV]& \multicolumn{2}{|c|}{$ 9.31\times10^7$} & \multicolumn{2}{|c}{$3.06\times10^6$}\\
$M_{N_3}$ [GeV]& \multicolumn{2}{|c|}{$2.19\times10^{14}$} & \multicolumn{2}{|c}{$2.02\times10^{13}$}\\
$\upsilon_R$ [GeV]& \multicolumn{2}{|c|}{$0.05\times 10^{16}$} & \multicolumn{2}{|c}{$0.18\times 10^{16}$}\\
\hline
\hline
\end{tabular}
\end{center}
\end{small}
\caption{Results from numerical fit corresponding to minimized $\chi^2$ for normal (NO)
and inverted ordering (IO) in neutrino masses. The fit is carried out for the GUT scale
extrapolated data given in Table \ref{GUT-data} for $\tan\beta=50$. The input parameters are
collected in Appendix.}
\label{fit-results}
\end{table}

The best fit obtained for IO and $\tan\beta=50$ is also shown in the Table \ref{fit-results}.  The
minimized value of $\chi^2$ is relatively large compared to the one obtained for NO but is
acceptable as all the observables are fitted within the $1\sigma$ range of their experimental
values. Note that the fitted profiles of the light neutrinos, {\it i.e.} $F_{\bar5} \sim {\tt
diag.}(\lambda^{0.8}, \lambda^{0.4},1)$, still follows the normal ordering structure (with a slightly
less hierarchical structure compared to that of NO case) while ${\cal O }(1)$ Yukawas in $Y_{126}$
conspire to create inverted ordering in the neutrino masses. The mismatch between the hierarchies in
neutrinos and charged fermions can be attributed more to a tuning of the ${\cal O}(1)$ Yukawa
couplings, rather than to an effect of the zero-mode profiles. We expect  that such a solution is
very sensitive to the Yukawa parameters in $Y_{126}$ and that even small deviations from their best
fit values can significantly raise the $\chi^2_{\rm min}$. We investigate this issue in the
following subsection.

\subsection{Anarchical  Yukawas: a test of naturalness}
So far the analysis implies that the model under consideration predicts approximate $t-b-\tau$
Yukawa unification which is compatible only with large values of $\tan\beta$. Moreover both the
normal and inverted ordering in the neutrino masses seem to be viable as indicated by the best fit
solutions. We do not know yet whether a successful fit in the two cases requires a special tuning of
the ${\cal O}(1)$ Yukawa parameters or not. In the present approach this question is relevant, since
the whole construction is based on the idea of anarchy in the Yukawa sector: the hierarchical
pattern of fermion masses and mixing angles is entirely due to the zero-mode profiles, while the
Yukawa couplings on the brane have no structure. If special relations between the ${\cal O}(1)$
Yukawa parameters were needed in order to reproduce the data, this would represent a fine-tuning
problem of our model, which cannot appeal to symmetry or dynamical principles to justify such
relations. If, on the contrary, the present model were natural, we would expect that a successful
explanation of fermion masses and mixing angles  should not  depend very much on the specific
choices of ${\cal O}(1)$ parameters. 

To investigate this feature, we have repeated the above analysis with some
changes. We randomly varied each of the complex elements in $Y_{10}$, $Y_{126}$ and $Y_{120}$ such
that $|Y_{ij}| \in [0.5,1.5]$ and ${\tt arg}(Y_{ij}) \in [0,2\pi]$, using flat distributions for
both.  For a given set of random Yukawa couplings, the $\chi^2$ is minimized versus the remaining $12$
parameters (4 bulk masses and 8 Higgs mixing parameters $\alpha_i$, $\bar\alpha_i$) using
the $17$ observables described earlier.  Unlike the previous case we investigate $10^5$ samples of random ${\cal O}(1)$ Yukawas and perform a $\chi^2$ minimization for each case. The analysis is carried out for $\tan\beta=50$ and for both NO and IO, 
as only these cases were found in good agreement with the GUT scale data
in the previous analysis, where also the Yukawa couplings were fitted.

\begin{figure}[!ht]
\centering
\subfigure{\includegraphics[width=0.5\textwidth]{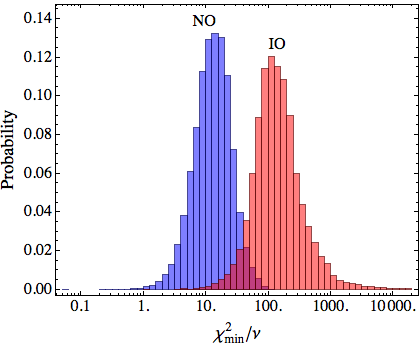}}
\caption{The distributions of minimized $\chi^2/\nu$  for NO and IO in neutrino masses and for
$\tan\beta=50$.}
\label{fig2}
\end{figure}

The results are displayed in Fig. \ref{fig2} where we show the distributions of
$\chi^2_{\rm min}/\nu$, $\nu=5$ being the number of independent degrees of freedom (d.o.f.) in the fit.
The distribution for NO is clearly peaked at lower values of $\chi^2_{\rm min}/\nu$, with respect 
to the one for IO. In Table \ref{tab3} we report the number of successful cases for different
threshold values of $\chi^2_{\rm min}/\nu$ together with the goodness of fit measured in terms of
$p-$values.
\begin{table}[!ht]
\begin{small}
\begin{center}
\begin{tabular}{lcccc}
 \hline
 \hline
~~~$p-$value & ~~~0.50~~~& ~~~0.10~~~ & ~~~{\bf 0.05}~~~& ~~~0.001~~~\\
$\chi^2_{\rm min}/\nu$ (for $\nu=5$)~~~~~ & $\le$ 0.87& $\le$ 1.85 & {\bf $\le$ 2.21} & $\le$ 4.10\\
\hline
successful cases (NO) & 0.1\% & 0.7\% & {\bf 1.2\%} & 5.6\%\\
successful cases (IO) & ~~$<10^{-3}$\%~~ & ~~$<10^{-3}$\%~~& ~~${\bf 10}^{\bf -3}${\bf \%}~~ &
0.01\%\\
\hline
\hline
\end{tabular}
\end{center}
\end{small}
\caption{The fraction of successful events obtained for different $p-$values from random samples of
${\cal O}(1)$ Yukawa couplings in case of normal and inverted ordering in the neutrino masses.}
\label{tab3}
\end{table}
The threshold $p \ge 0.05$ is often considered as an acceptable value for the statistical validity of a
fit. As it can be seen from Table \ref{tab3}, in the NO case $p$ is larger than
$0.05$ in about one percent of the generated samples, while in the IO case only one over total
$10^5$ samples reaches the modest $p$-value of $0.05$. Hence the NO turns out as a more natural
choice in this model. 

In the NO case one percent can be regarded as the size of the required tuning to reproduce
the data within the framework of anarchy. It is clear that this number has no absolute meaning
and could only be useful if compared with analogous numbers obtained by analyzing other models with
a similar approach. A success rate of order $0.01$ is a typical outcome in this kind of analysis
for the most successful models \cite{Altarelli:2002sg,Buchmuller:2011tm,*Bergstrom:2014owa}.

\begin{figure}[!ht]
\centering
\subfigure{\includegraphics[width=0.4\textwidth]{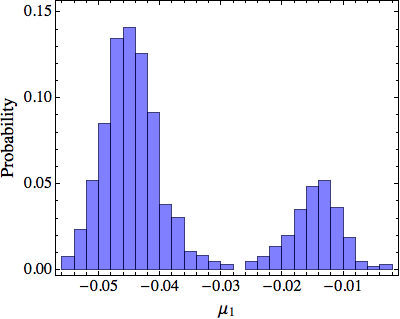}}\quad
\subfigure{\includegraphics[width=0.41\textwidth]{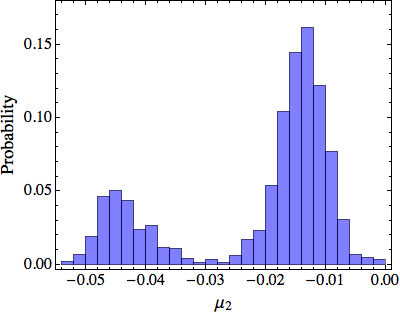}}\\
\subfigure{\includegraphics[width=0.4\textwidth]{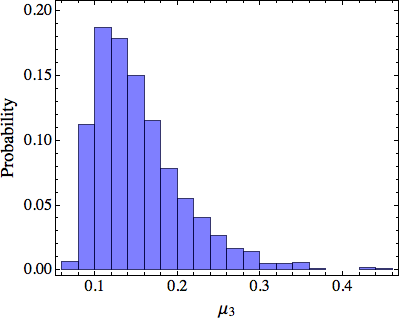}}\quad
\subfigure{\includegraphics[width=0.4\textwidth]{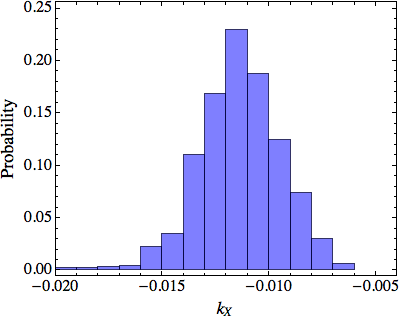}}
\caption{The distributions of bulk mass parameters fitted with $\chi^2_{\rm min}/\nu<2.21$ in case
of NO and
$\tan\beta=50$.}
\label{fig3}
\end{figure}

\begin{figure}[!ht]
\centering
\subfigure{\includegraphics[width=0.325\textwidth]{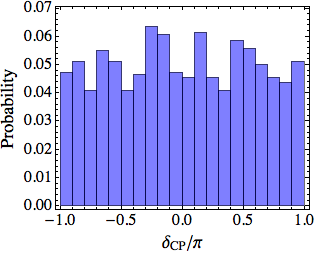}}\quad \hspace*{-0.25cm}
\subfigure{\includegraphics[width=0.31\textwidth]{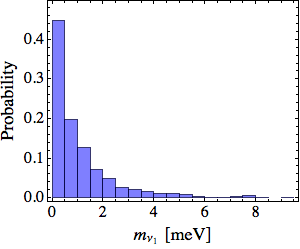}}\quad \hspace*{-0.25cm}
\subfigure{\includegraphics[width=0.32\textwidth]{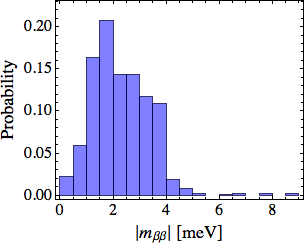}}\quad
\caption{The predictions for various observables obtained for $\chi^2_{\rm min}/\nu<2.21$ in case of
NO and
$\tan\beta=50$.}
\label{fig4}
\end{figure}

 The probability distributions for the bulk mass parameters obtained in 1.2\% of the NO cases
corresponding to $p \ge 0.05$  are shown in Fig. \ref{fig3}.  One finds $\mu_3>\mu_2\ge\mu_1$ in most of the cases as expected and $k_X$ turns out to be ${\cal O}(\mu_{1,2})$. A few cases described by smaller peaks in the distributions of $\mu_1$ and 
$\mu_2$  corresponds to $\mu_1 > \mu_2$. However, such cases are equivalent to the cases with
$\mu_1 < \mu_2$ as one can always interchange  ${\bf 16}_1 \leftrightarrow {\bf 16}_2$ by
interchanging $\mu_1$ and $\mu_2$ and also the first and second rows and columns of all the Yukawa
coupling matrices on the brane. Such a transformation on Yukawa matrices still preserves their
anarchical structure and both these pictures lead to the same physical scenario. As it can be
seen from Fig. \ref{fig3}, the preferred values of all the bulk masses remain well below the cut-off
scale, and they do not endanger the validity of the effective theory. The $k_X< 0$ leads to a
relatively weak hierarchy among the down-type quarks and charged leptons in comparison to that in
the up-type quarks. From the most probable values of $\mu_i$ and $k_X$ of Fig. \ref{fig3} we get, 
\be \label{prob-profiles}
F_{10} \simeq \lambda^{0.4}~{\tt diag}(\lambda^{4.1}, \lambda^{2.2},1)~~{\rm and}~~F_{\bar 5}
\simeq \lambda^{0.3}~{\tt diag}(\lambda^{1.5}, \lambda^{0.7},1).
\ee
The above profiles of zero modes provide a quantitative understanding of the differences between
the quarks and lepton masses and mixing patterns.
The successful cases corresponding to $\chi^2_{\rm min}/\nu<2.21$ can also be used to derive the
predictions for the other observables in the lepton sector. The probability distributions for the
leptonic Dirac CP phase, the lightest neutrino mass and the effective mass of neutrinoless double
beta decay $|m_{\beta \beta}|$ are shown in Fig. \ref{fig4}.
 One finds an almost uniform distribution in $\delta^l_{CP}$ and the entire range 
in CP phase is allowed by the model. The lightest neutrino mass is restricted to be  $\lesssim 5$
meV corresponding to a hierarchical neutrino mass spectrum while $|m_{\beta\beta}|$ is predicted in
the range 0.1-5 meV. Both these predictions are far from the sensitivity of current generation
experiments and any positive signal in these experiments would essentially rule out the model.
 
\begin{figure}[!ht]
\centering
\subfigure{\includegraphics[width=0.45\textwidth]{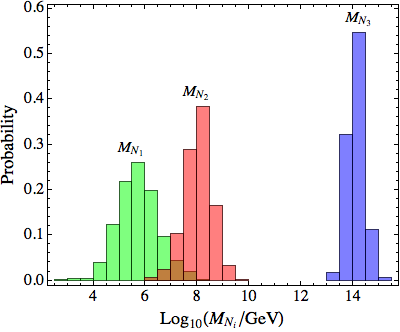}}\quad
\subfigure{\includegraphics[width=0.46\textwidth]{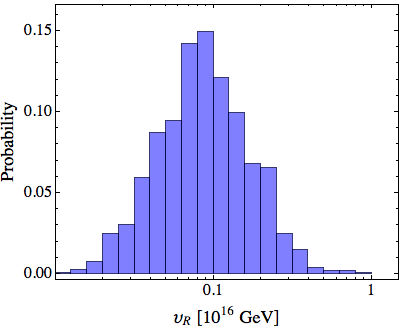}}\quad
\caption{The predictions for the masses of RH neutrinos and the VEV of ${\overline{\bf 126}_H}$
obtained for $\chi^2_{\rm min}/\nu<2.21$ in case of NO and $\tan\beta=50$.}
\label{fig5}
\end{figure}

The predictions for the RH neutrino masses and the VEV of ${\overline{\bf 126}_H}$ are displayed in
Fig. \ref{fig5}. The bulk masses of the singlets in ${\bf 16}_i$ are predicted from the fitted values
of $\mu_i$ and $k_X$. From their most probable values, we obtain
\be 
\label{prob-RHN-profile}
F_1\simeq \lambda^{0.6}~{\tt diag} (\lambda^{7.0}, \lambda^{5.0}, 1)~.
\ee
This results into an extremely hierarchical mass spectrum for the RH neutrinos corresponding to
$M_{N_1}\approx \lambda^{15} \upsilon_R$, $M_{N_2}\approx \lambda^{11}
\upsilon_R$ and $M_{N_3}\approx \lambda \upsilon_R$, as shown in the left panel of Fig. \ref{fig5}. As
explained earlier, the masses of the RH neutrinos do not play any role in the seesaw mechanism but
they can be important for leptogenesis. For a hierarchical mass spectrum of RH neutrinos, a
successful thermal leptogenesis  requires the mass of the lightest RH neutrino $M_{N_1} \gtrsim
10^9$ GeV \cite{Davidson:2002qv,*Buchmuller:2002rq} in a standard flavour independent scheme.  When
flavour effects are considered, it is possible to generate a sufficient lepton asymmetry through the
decay of the next-to-lightest neutrinos if $10^{12}~{\rm GeV} \gtrsim M_{N_2}\gtrsim10^9~{\rm GeV}$
and $M_{N_1}\ll 10^9$ GeV as suggested in \cite{DiBari:2005st,*Bertuzzo:2010et,*DiBari:2013qja}.
Both these alternatives cannot be realized in this model, which predicts $M_{N_1} \ll M_{N_2} <
10^9$ GeV.

The scale of atmospheric neutrino oscillation requires the VEV of ${\overline{\bf 126}_H}$ at
least an order of magnitude below $M_{\rm GUT}$, as it can be seen from Fig. \ref{fig5}. This
is compatible with the D-term cancellation condition, Eq. (\ref{Dterm-cond}). As can
be seen from Fig. \ref{fig3}, the viable fits to the fermion masses and mixing angles require
$\upsilon_\Phi^{3/2}=k_X \Lambda/(\sqrt{2}g_5) < 0$. Hence the VEV of ${\bf 126}_H$ can cancel the
D-term in Eq. (\ref{Dterm-cond}) even if $|\langle{\overline{\bf 126}_H}\rangle| \ll |\langle{\bf
126}_H\rangle| \sim M_{\rm GUT}$.

\begin{figure}[!ht]
\centering
\subfigure{\includegraphics[width=0.45\textwidth]{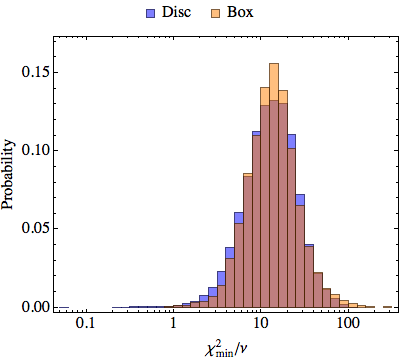}}\quad
\subfigure{\includegraphics[width=0.45\textwidth]{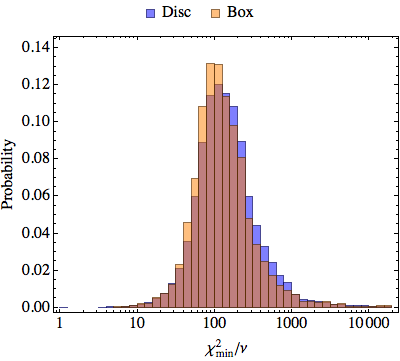}}\quad
\caption{A comparison between the minimized $\chi^2$ distributions obtained from the two different
 distribution of random ${\cal O}(1)$ Yukawa couplings (see text for the details). The left (right)
panel corresponds
to the NO (IO) case and $\tan\beta = 50$.}
\label{fig6}
\end{figure}

We conclude this section with a comment on the choice of the ${\cal O}(1)$ Yukawa parameters. For the
above analysis, we have randomly chosen them from a flat distribution of points residing on the disc
of inner radius 0.5 and outer radius 1.5 in a complex plane. To assess the dependence of our results
on the criteria for selecting the Yukawa parameters, we have repeated a similar
analysis for random Yukawa couplings residing in the box of vertices $(1+i, -1+i, -1-i, 1-i)$ in a complex
plane. The results are shown in Fig. \ref{fig6} where we compare the probability distributions
obtained for the two choices of Yukawa parameters. The obtained distributions are almost
indistinguishable and the new choice leads to nearly the same results as the old one. The overall results
and predictions derived in this subsection are therefore robust and should stand for similar choices
of ${\cal O}(1)$ parameters.

\section{Conclusions}
\label{conclusions}
Anarchical Yukawa matrices for both quarks and leptons can be nicely reconciled with the observed 
fermion masses and mixing angles by appealing to wave-function renormalizations that distinguish
generations and fermion species. This naturally occurs in models with an extra spatial dimension,
thanks to the different localization of the profiles of fermion zero modes. Such a picture is also
compatible with SU(5) grand unification, at least in first approximation. The large mixing angles
of the lepton sector and the moderate hierarchy among neutrino masses is attributed to a nearly
equal wave function renormalization of the three generations of $\bar 5$. The observed hierarchies
in the charged lepton sector and in the quark sectors are mostly due to the different
renormalization affecting the three generations of $10$. The different rescaling of $\bar 5$-plets
and 10-plets implies that mass ratios in the up-quark sector are nearly the square of the
corresponding mass ratios in the down-quark and charged-lepton sectors, which is approximately true.
Such a description can be extended to SO(10), as shown in a five-dimensional model by Kitano and
Li. Before SO(10) breaking, all members of a fermion generation, hosted in a ${\bf 16}$
representation, have the same zero-mode profile, which depends upon an SO(10) invariant bulk mass
term. In the Kitano-Li model SO(10) is broken down to the direct product of SU(5) and U(1)$_{\rm
X}$, by the VEV of an SO(10) adjoint that lives in the bulk and has gauge coupling to matter
supermultiplets. In this way different bulk mass terms for the SU(5) multiplets residing in the
same ${\bf 16}$ representation are generated, with corresponding different zero-mode profiles.

In this paper we have reconsidered the Kitano-Li model. To start with, we have modified it in such a
way that fermion masses and mixing angles are dominated by Yukawa interaction terms of the same
dimensionality, while in the original model non-renormalizable operators of different mass
dimensions had to provide comparable contributions to achieve a realistic description. We have also
included a set of Higgs multiplets that allows for a solution to the doublet-triplet splitting
problem through the missing partner mechanism. We have explicitly specified the set of interactions
needed in order to implement such a mechanism. Finally, we have tested the validity of the model by
realizing a series of fits to an idealized set of 17 data, obtained by naively extrapolating fermion
masses and mixing angles from low energy to the GUT scale. Our model depends on 27 anarchical Yukawa
couplings, 8 parameters characterizing the light Higgs combinations and 4 parameters that describe
the fermion bulk masses. In a first fit we left all parameters to vary freely and we obtained an excellent agreement with data for
both the cases of normal and inverted ordered neutrino spectrum, for large values of $\tan\beta$.
Despite the large number of free parameters, we consider such an agreement not completely trivial,
given the fact that 35 of our parameters can vary in a very limited interval close to one and that
only the 4 parameters describing the bulk masses are responsible for all the observed hierarchies
in the fermion spectrum. In a second stage, to detect a possible fine-tuning among the anarchical 
Yukawa couplings, we have modified our numerical analysis by first generating a random sample of
${\cal O}(1)$ Yukawa couplings and by subsequently fitting the remaining 12 Higgs and
bulk parameters. This procedure have been iterated to obtain a distribution of minimum $\chi^2$
values. We see a clear difference between the cases of normal ordering and inverted ordering in the
neutrino masses. While in the inverted ordering case we need about $10^5$ samples to reach a
$p$-value close to $0.05$, in the normal hierarchy case in about one percent of the cases we have
$p>0.05$. This unambiguously indicates that our model needs a severe fine-tuning of the
``anarchical'' parameters in the case of inverted ordered neutrino spectrum
while the normally ordered one is accommodated much more naturally.

We verified that these results are stable versus changes in the drawing of the anarchical
parameters. Our analysis is incomplete under several respects. In particular we have not discussed
the breaking of $N=1$ supersymmetry. This would have allowed to reduce the uncertainty in the
extrapolation of fermion masses and mixing angles from low-energy to the GUT scale, at the cost of a
much bigger model dependence. Consequently we were not able to analyze the rich related
phenomenology of flavour and CP violations, both in the quark and in the lepton sector, which
heavily depends on the chosen mechanism for supersymmetry breaking.

 Concerning predictions, we have found no preference for any particular value of the 
leptonic Dirac CP phase. The
lightest neutrino mass should lie below $5$ meV, corresponding to a hierarchical neutrino mass
spectrum while $|m_{\beta\beta}|$ is predicted in the range 0.1-5 meV. Any positive signal in the
current generation of experiments aiming at measuring neutrino masses or $|m_{\beta\beta}|$ in the
lab would essentially rule out the model. The hierarchy in the right handed neutrino spectrum is
very pronounced and the corresponding mass distributions are peaked around $10^6$ GeV, $10^8$ GeV
and $10^{14}$ GeV. As a consequence, thermal leptogenesis cannot be responsible for the observed
baryon asymmetry in our model. On a more qualitative side, our analysis confirms that the idea of
anarchical Yukawa couplings can be successfully implemented even in the context of an SO(10) grand
unified theory. The required conditions are that SO(10) is broken down to SU(5) as a first step and
that the anarchical Yukawa couplings of the different charge sectors are not entirely dominated by a
single SO(10) Higgs multiplet. Special features of neutrino data, such as the indications of a
nearly maximal atmospheric mixing angle and hierarchical light neutrino spectrum  should be regarded as
fully accidental in this approach, devoid of any implications on the underlying dynamics.

\begin{acknowledgments}
We thank Stefano Rigolin for participating to the first stage of the present investigation. We thank
Pasquale Di Bari for stimulating discussions on leptogenesis in the context of the model presented
in this article. We acknowledge partial support from the European Union network FP7 ITN INVISIBLES
(Marie Curie Actions, PITN-GA-2011-289442). K.M.P. thanks the Department of Physics and Astronomy of
the University of Padova for its support.
 \end{acknowledgments}

\appendix
\section{Parameters obtained for the best fit solutions}
We give here the input  parameters obtained for the best fit solutions presented in the Table \ref{fit-results} corresponding  to normal and inverted  neutrino mass spectrum  and $\tan\beta = 50$. 

\subsection{Normal ordering}
The best fit values of the Yukawa matrices and various parameters appearing in Eqs. 
(\ref{Yukawa-sum-rule}, \ref{seesaw}) at $\chi^2_{\rm min}\approx 0$ are listed below.
\beqa 
Y_{10}&=&
\left(
\begin{array}{ccc}
 0.78314 & 1.05610 e^{1.79809 i}& 0.92306 e^{-0.19874 i}\\
 1.05610 e^{ 1.79809 i}& 1.49012 & 1.09077 e^{1.02405 i}\\
 0.92306 e^{ -0.19874 i}& 1.09077 e^{1.02405 i}& 0.96156
\end{array}
\right)
~, \nonumber \\
Y_{120}&=&\left(
\begin{array}{ccc}
 0 & 1.04750 e^{-2.22311 i} & 0.50164 e^{3.02587i} \\
 -1.04750 e^{-2.22311 i}& 0 & 0.78048 e^{-0.43312 i}\\
 -0.50164 e^{3.02587i}& -0.78048 e^{-0.43312 i}& 0
\end{array}
\right)
~, \nonumber \\
Y_{126}&=&\left(
\begin{array}{ccc}
 1.49976 e^{ 1.45362i} & 0.51701  e^{-1.20768 i}& 1.25349  e^{-1.82494 i}\\
 0.51701  e^{-1.20768 i}& 0.50067 e^{-0.49914i} & 0.91593  e^{0.05375 i}\\
 1.25349 e^{-1.82494 i} & 0.91593  e^{0.05375 i}& 1.38243 e^{-1.37536 i}
\end{array}
\right)~. \eeqa
\beqa
\alpha_2&=&0.87373~ e^{-2.05269 i}~,~~\alpha_3=0.06975~ e^{-1.21499 i}~,\nonumber \\
\overline{\alpha}_2&=&0.81115~ e^{-1.00683 i}~,~~\overline{\alpha}_3=0.50212~ e^{-1.20195 i}~.
\eeqa
The bulk mass parameters are:
\beqa
\{\mu_1,~\mu_2,~\mu_3,~k_X\} &=& \{-0.03732,~-0.01565,~0.20467,~-0.01031\}~.\eeqa

\begin{figure}[!ht]
  \centering
 \subfigure{\includegraphics[width=0.31\textwidth]{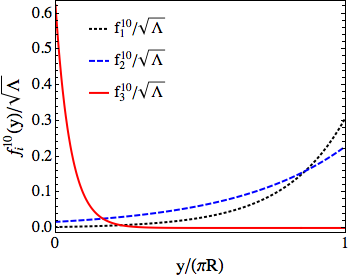}}\quad
 \subfigure{\includegraphics[width=0.31\textwidth]{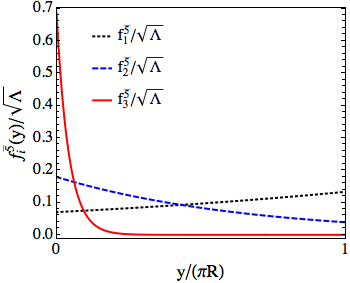}}\quad
 \subfigure{\includegraphics[width=0.31\textwidth]{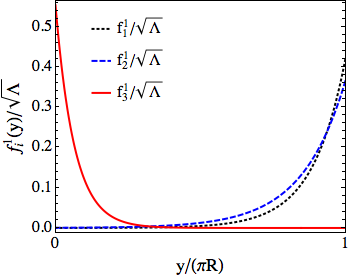}}
 \caption{The localized zero-mode profiles of different SU(5) matter multiplets for the best fit 
 solution obtained in case of  normal ordering in the neutrino masses.}
 \label{fig-profile-NO}
 \end{figure}

From the above parameters and using Eq. (\ref{y5}), one obtains the following zero-mode profiles 
of various SU(5) multiplets at $y=0$.
\beqa
F_{10}=\lambda ^{0.3} \left(
\begin{array}{ccc}
 \lambda ^{3.7} & 0 & 0 \\
 0 & \lambda ^{2.4} & 0 \\
 0 & 0 & 1
\end{array}
\right),~
F_{\bar5} = \lambda ^{0.3} \left(
\begin{array}{ccc}
 \lambda ^{1.5} & 0 & 0 \\
 0 & \lambda ^{0.9} & 0 \\
 0 & 0 & 1
\end{array}
\right),~
F_1=  \lambda ^{0.4} \left(
\begin{array}{ccc}
 \lambda ^{6.2} & 0 & 0 \\
 0 & \lambda ^{4.8} & 0 \\
 0 & 0 & 1
\end{array}
\right).\eeqa
Further, the localization of zero-mode profiles of different generations of 10, ${\bar 5}$ and 1 
can be obtained by replacing $m \rightarrow m_i^r$ in Eq. (\ref{zero-mode}) and are displayed in
Fig. \ref{fig-profile-NO}. 

\subsection{Inverted ordering}
The best fit values of the Yukawa matrices and various parameters appearing in Eqs.
(\ref{Yukawa-sum-rule}, \ref{seesaw}) at $\chi^2_{\rm min}\approx 5.75$ are listed below.
\beqa  
Y_{10}&=&\left(
\begin{array}{ccc}
 0.50232  & 1.12746 e^{ -0.07927 i}& 1.25370 e^{-0.46501  i}\\
 1.12746 e^{ -0.07927 i}& 0.74605 & 1.49999 e^{ 0.94538 i}\\
 1.25370 e^{ -0.46501 i}& 1.49999 e^{ 0.94538 i}& 1.38633 
\end{array}
\right)~, \nonumber \\
Y_{120}&=&\left(
\begin{array}{ccc}
 0 & 0.53486 e^{2.05559 i} & 1.29570 e^{2.53388i} \\
 -0.53486 e^{2.05559 i} & 0 & 0.58945 e^{-0.03658 i} \\
 -1.29570 e^{2.53388 i} & -0.58945 e^{-0.03658 i} & 0
\end{array}
\right)~, \nonumber \\
Y_{126}&=&\left(
\begin{array}{ccc}
 1.49999  e^{0.16531  i}& 0.50005  e^{1.43459  i}& 0.58661  e^{-2.24612 i}\\
 0.50005  e^{1.43459 i}& 0.50007  e^{-0.86236 i}& 1.02973  e^{-1.58869 i}\\
 0.58661  e^{-2.24612 i}& 1.02973  e^{-1.58869 i}& 0.96577 e^{-0.16857 i}
\end{array}
\right)~. \eeqa
\beqa
\alpha_2 &=& 0.04681~e^{1.46923 i}~,~~\alpha_3=0.07100 ~e^{3.10679 i}~,\nonumber \\
\overline{\alpha}_2&=&0.87191~e^{0.46323 i}~,~~\overline{\alpha}_3=0.36100~ e^{-1.36707 i}~.
\eeqa 
The bulk mass parameters  are:
\beqa
\{\mu_1,~\mu_2,~\mu_3,~k_X\} &=& \{-0.040351,~-0.01099,~0.085029,~-0.01668\}~.\eeqa

\begin{figure}[!ht]
  \centering
 \subfigure{\includegraphics[width=0.315\textwidth]{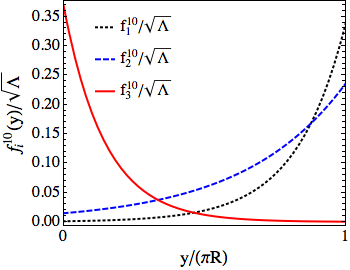}}\quad
 \subfigure{\includegraphics[width=0.31\textwidth]{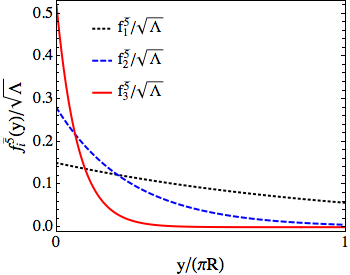}}\quad
 \subfigure{\includegraphics[width=0.31\textwidth]{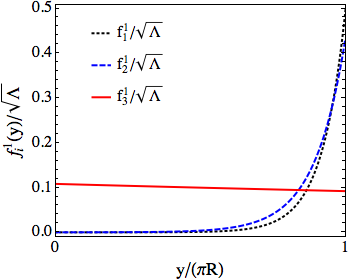}}
 \caption{The localized zero-mode profiles of different SU(5) matter multiplets for the best fit 
 solution obtained in case of inverted ordering in the neutrino masses.}
 \label{fig-profile-IO}
 \end{figure}

From the above parameters and using Eq. (\ref{y5}), one obtains the following zero-mode profiles  
of various SU(5) multiplets at $y=0$.
\beqa
F_{10}= \lambda ^{0.7} \left(
\begin{array}{ccc}
 \lambda ^{3.9} & 0 & 0 \\
 0 & \lambda ^{2.2} & 0 \\
 0 & 0 & 1
\end{array}
\right),~
F_{\bar5} = \lambda ^{0.4} \left(
\begin{array}{ccc}
 \lambda ^{0.8} & 0 & 0 \\
 0 & \lambda ^{0.4} & 0 \\
 0 & 0 & 1
\end{array}
\right),~
F_1=  \lambda ^{1.5} \left(
\begin{array}{ccc}
 \lambda ^{7.4} & 0 & 0 \\
 0 & \lambda ^{5.5} & 0 \\
 0 & 0 & 1
\end{array}
\right).\eeqa
 Further, the localization of zero-mode profiles of different generations of 10, ${\bar 5}$ and 1
 fermions  can be obtained by replacing $m \rightarrow m_i^r$ in Eq. (\ref{zero-mode}) and are
displayed in Fig. \ref{fig-profile-IO}. 

\bibliographystyle{apsrev4-1}
\bibliography{ref-so10.bib}
\end{document}